\documentclass[twocolumn,prd,superscriptaddress,nofootinbib]{revtex4-1}

\usepackage{amsfonts,amsmath,amssymb,mathrsfs}
\usepackage{hyperref}
\usepackage{color}
\usepackage{graphicx}  % needed for figures
\usepackage{dcolumn}   % needed for some tables
\usepackage{bm}        % for math
\usepackage[english]{babel}
\usepackage[bottom]{footmisc}
\usepackage{dblfnote}
\usepackage{mathtools}
\usepackage{amsmath}
\usepackage{physics}
\DFNalwaysdouble % for this example

\def\a{\alpha}

\def\r{\rho}
\def\s{\sigma}
\def\t{\tau}
\def\m{\mu}
\def\n{\nu}
\def\k{\kappa}
\def\th{\theta}
\def\g{\gamma}\def\G{\Gamma}
\def\L{t}\def\l{V}
\def\D{\Delta}
\def\la{\langle}
\def\ra{\rangle}
\def\o{\omega}\def\O{\Omega}
\def\d{\delta}
\def\p{\partial}

\def\oxthree{{\cal O}(x^3) }

\def\half{\textstyle{\frac{1}{2}}}

\def\bdoc{\begin{document}}
\def\edoc{\end{document}}
\def\bea{\begin{equation}}
\def\eea{\end{equation}}

\def\beq{\begin{eqnarray}}
\def\eeq{\end{eqnarray}}
\def\be{\begin{eqnarray}}
\def\ee{\end{eqnarray}}
\def\ben{\begin{enumerate}}
\def\een{\end{enumerate}}
\def\la{\langle}
\def\ra{\rangle}
\def\a{\alpha}
\def\g{\gamma}\def\G{\Gamma}
\def\d{\delta}\def\D{\Delta}
\def\e{\epsilon}
\def\z{\zeta}

\def\th{\theta}
\def\k{\kappa}
\def\l{t}
\def\m{\mu}
\def\n{\nu}
\def\o{\omega}
\def\p{\pi}
\def\r{\rho}
\def\s{\sigma}
\def\t{\tau}
\def\L{{\cal L}}
\def\S{\Sigma }
\def\gsim{\; \raisebox{-.8ex}{$\stackrel{\textstyle >}{\sim}$}\;}
\def\lsim{\; \raisebox{-.8ex}{$\stackrel{\textstyle <}{\sim}$}\;}
\def\gtrsim{\gsim}
\def\lessim{\lsim}
\def\loc{{\rm local}}
\def\vm{v_{\rm max}}
\def\bh{\bar{h}}
\def\del{\partial}
\def\nab{\nabla}
\def\half{{\textstyle{\frac{1}{2}}}}
\def\fourth{{\textstyle{\frac{1}{4}}}}

\def\bD{{\bf D}}
\def\bE{{\bf E}}
\def\bF{{\bf F}}
\def\bB{{\bf B}}
\def\bP{{\bf P}}
\def\bV{{\bf v}}
\def\bv{{\bf v}}
\def\bx{{\bf x}}
\def\by{{\bf y}}
\def\bz{{\bf z}}
\def\ba{{\bf a}}
\def\bd{{\bf d}}
\def\bs{{\bf s}}
\def\bn{{\bf n}}
\def\bp{{\bf p}}

\def\O{\Omega}

\def\br{{\bf r}}
\def\bnab{{\bf \nab}}

\def\tE{\tilde{E}}
\def\tL{\tilde{L}}
\def\Horava{Ho\v{r}ava }

\def\oxtwo{\mathscr{O}\left(x^2\right)}
\def\oxthree{\mathscr{O}\left(x^3\right)}
\def\oxfour{\mathscr{O}\left(x^4\right)}
\def\oxfive{\mathscr{O}\left(x^5\right)}
\def\LL{\text{Lanczos-Lovelock}}

\def\ph{\phantom}

\begin{document}
\title{
Hawking radiation from stationary black holes using gravitational anomaly }
\author{Selim Sk}
\email{selimsk@iitgn.ac.in }
\affiliation{Indian Institute of Technology, Gandhinagar, Gujarat 382055, India.}
\author{Sudipta Sarkar}
\email{sudiptas@iitgn.ac.in}
\affiliation{Indian Institute of Technology, Gandhinagar, Gujarat 382055, India.}

%\date
\begin{abstract}
Among all the different techniques to derive the Hawking effect, the approach based on gravitational anomaly by Robinson and Wilczek provides a simple and satisfactory origin of the black hole radiation. In this picture, the effective near horizon physics becomes chiral and contains gravitational anomaly. Nevertheless, the underlying description must be generally covariant, and therefore we require a compensating energy-momentum flux whose divergence cancels the anomaly at the horizon. Remarkably, the energy flux associated with the Hawking emission from the horizon exactly cancels the gravitational anomaly and restores the general covariance at the quantum level. In this work, we present a generalization of the original derivation for a stationary axisymmetric black hole solution of any gravity theory which differs perturbatively from general relativity. The crucial input of the calculation is a remarkable simplification of the near horizon geometry and the validity of the zeroth law of black hole mechanics.

 \end{abstract}
\maketitle
\section{Introduction}
The physics of black holes has provided remarkable insights into the nature of quantum theory in curved spacetime. The most intriguing prediction of the quantum theory in the presence of a black hole is the existence of Hawking radiation. In the original work by Hawking \cite{Hawking:1975vcx}, the origin of the black body radiation lies in the time dependence of the background collapsing geometry, which populates the late time vacuum by Hawking quanta. The black hole temperature is calculated by evaluating the Bogolyubov coefficients between the asymptotic `in' and `out' vacuum states. The result is a Planckian spectrum at temperature $ T = \hbar \kappa/ 2\pi $, where $\kappa$ is the surface gravity associated with the final stationary black hole.\\

Since the original derivation by Hawking, the same result has been derived from various approaches. Apart from the canonical derivation, a path-integral approach can be used to obtain the same result \cite{Hartle:1976tp}. Also, the radiation can be ascribed to the tunneling of virtual particles from an eternal black hole \cite{Parikh:1999mf}. Since all these approaches yield identical results, the Hawking effect is considered a solid prediction of quantum field theory in curved spacetime. In fact, the derivation of Hawking radiation is also regarded as a low energy consistency check of any proposal for quantum gravity.\\

Among all the different techniques to derive the Hawking effect, the approach based on gravitational anomaly \cite{Robinson:2005pd} provides a simple and satisfactory origin of the black hole radiation. In this picture, the effective near horizon physics becomes chiral and contains gravitational anomaly. Nevertheless, the underlying description must be generally covariant, and therefore we require a compensating energy-momentum flux whose divergence cancels the anomaly at the horizon. Remarkably, the energy flux associated with the Hawking emission from the horizon exactly cancels the gravitational anomaly and thereby restores the general covariance of the theory at the quantum level. Therefore, the Hawking flux originated from the covariance of quantum field theory near the horizon of a black hole. \\

The original derivation of the Hawking radiation from anomaly was done for a spherically symmetric black hole spacetime. The derivation is later generalized to stationary Kerr and time-dependent Vaidya black holes \cite{Murata:2006pt, Vagenas:2006qb}. In all cases, the derivation depends on the crucial fact that the effective physics near the horizon is essentially $(1+1)$ dimensional; the only relevant part is the $`r - t$' sector of the metric.\\

In this work, we generalize the derivation of the Hawking radiation from gravitational anomaly beyond spherical symmetry and for general static as well as stationary black holes. We show that the derivation is possible because of a remarkable simplification of the near-horizon geometry, which is demonstrated by \cite{Medved:2004ih, Medved:2004tp}. As in the spherically symmetric case, close to the black hole horizon, the field theory can be described using an infinite collection of (1 + 1)-dimensional fields, each propagating in a spacetime with a  two-dimensional metric. This allows us to repeat the calculation of the anomaly canceling flux as the Hawking radiation from the horizon. Our calculation also indicates an interesting relationship between the derivation of the Hawking flux and the applicability of the zeroth law of black hole mechanics.

\section{Static Black Hole: Geometric Set up}

In this section, we will present a generalization of the derivation of Hawking radiation using gravitational anomaly for a static black hole. Before proceeding with the main derivation, we recall that an anomaly in a quantum field theory is a conflict between a symmetry of the classical action and the procedure of quantization. While the anomaly associated with global symmetries indicates interesting physics, the gauge anomaly, i.e., the violation of a local gauge invariance by quantum effects, signals theoretical pathology. All gauge anomalies must cancel out; otherwise, it will lead to an inconsistency in the quantum theory, particularly unphysical negative norm states. \\

In classical gravity, the conservation of a matter energy-momentum tensor $T_{\mu \nu}$ is due to the diffeomorphism invariance of the theory. This can easily be verified by calculating the variation of the action functional of gravity under boundary preserving diffeomorphism. Nevertheless, quantum effects may lead to the non-conservation of $T_{\mu \nu}$ leading to the violation of the general covariance. This gravitational anomaly occurs when Weyl fermions or self-dual anti-symmetric tensor fields are coupled to gravity \cite{Alvarez-Gaume:1983ihn}. Also, as a simple model, gravitational anomaly arises when we consider a chiral scalar field in $1+1$ dimensions. The expression of the anomaly is then given by \cite{Bertlmann:2000da, Bertlmann:1996xk}

\begin{equation}\label{anomaly}
    \nabla_\mu T^{\mu}_{\nu} = \frac{1}{96 \pi \sqrt{-g}} \epsilon^{\beta \delta} \partial_\delta \partial_\alpha \Gamma^{\alpha}_{\nu \beta}.
\end{equation}

In \cite{Robinson:2005pd}, the notion of gravitational anomaly is used to model the near horizon physics and to derive the Hawking radiation. The key idea is to consider the near horizon effective theory by tracing over modes which leads to a singular contribution of the energy-momentum tensor. The proposal was to remove these modes from the theory near the black hole horizon at the expense of making the theory chiral. This is exactly similar to the setup used in \cite{Fulling:1986rk} to derive the gravitational anomaly for a chiral scalar field. This gravitational anomaly presents a serious inconsistency in this prescription, leading to a violation of general covariance. Then, it turns out that the gravitational anomaly in the form of a current can be exactly canceled by the flux of Hawking radiation. Therefore, in this picture, Hawking radiation arises as a compensating current canceling the troublesome gravitational anomaly and restoring the general covariance of the near horizon physics.\\

The crucial mathematical step which allows us to complete the derivation is the fact that the action of a scalar field in a $D$-dimensional spacetime in the near horizon region can be described using an infinite collection of $(1 + 1)$ -dimensional fields, such that the effective near horizon physics is only $1+1$ dimensional and we can use the result of \cite{Robinson:2005pd}. This step requires explicit use of the spherical symmetry of the problem. Later, it was generalized to Kerr black holes also \cite{Murata:2006pt}. Nevertheless, we do not have a demonstration of this simplification for general static and stationary black holes. In this work, we aim to achieve such a generalization.\\

We start with a static, asymptotically flat black hole spacetime described by the metric \cite{Medved:2004ih},

\begin{equation}\label{staticmetric}
ds^2=-N^2(n,x^a) dt^2+dn^2+\gamma_{ab}(n,x^a)dx^adx^b,
\end{equation}

\noindent
where $a=2,3\cdots$ . The existence of a Killing vector field, $(\partial_t)^\mu$, for the metric in Eq.(\ref{staticmetric}) indicates that spacetime has time-translational symmetry. We are using the Gaussian coordinate system $(t, n, x^a)$ with $n$ denoting the normal distance to the horizon, and $x^a$ are arbitrary coordinates on the (D-2) spacelike surface. Then the norm of the timelike Killing vector vanishes at $N = 0$, and that is the location of the Killing horizon of the spacetime. We chose the coordinates such that $n=0$ implies $N(n,x^a) = 0$. Then, the surface gravity associated with this Killing horizon is defined as

\begin{equation}
    \kappa \equiv \lim_{n\rightarrow 0}\partial_nN.
\end{equation}

Though it is not obvious, the surface gravity $\kappa$ is actually a constant on the horizon of a static black hole, independent of the transverse coordinates $x^a$. We will extensively use this property in the derivation.\\

In this geometry, consider a minimally coupled real massless scalar field $\varphi(x^\mu)$, described by the action

\begin{equation}
  S[\varphi]=\frac{1}{2}\displaystyle{\int d^Dx\sqrt{-g}\,\varphi\hspace{0.5mm} \Box\varphi}.  
\end{equation}

For the static background, described by the metric in Eq. (\ref{staticmetric}), the action for the scalar field becomes
\begin{align}\label{action}
S[\varphi]&=\frac{1}{2}\int\, d^Dx\sqrt{\gamma}\,\varphi \Big[-\frac{1}{N}\partial^2_t\varphi+\frac{1}{\sqrt{\gamma}}\partial_n\Big(N\sqrt{\gamma}\partial_n\varphi\Big) \nonumber\\&+\frac{1}{\sqrt{\gamma}}\Big\{N\partial_a\Big(\sqrt{\gamma} \gamma^{ab}\partial_b\varphi\Big)\nonumber \\&+\partial_aN\Big(\gamma^{ab}\sqrt{\gamma}\Big)\partial_b\varphi\Big\}\Big]. 
\end{align}
where $ \gamma$ is the determinant of the metric $\gamma_{ab}$. We like to know the form of this action in the near horizon limit.  It is worth noting that in the horizon limit, the second term in the Eq.(\ref{action}) becomes $\partial_n(N\partial_n\varphi)$, and the third term vanishes due to $N(n,x^a)=0$ at the horizon. In order to make the theory effectively two-dimensional near the horizon, we must eliminate the transverse coordinate dependence in Eq.(\ref{action}). As a result, it is necessary for us to express the lapse function as $N(n,x^a)= f(n)G(x^a)$ only in the vicinity of the horizon with the property at $n\rightarrow 0$ limit $f(n)=0$. Due to this expression of the lapse function, the fourth term in Eq.(\ref{action}) becomes zero in the horizon limit, and the action, $S[\varphi]$ takes the following simple form,
\begin{align}\label{actionNH}
S[\varphi]&=\frac{1}{2}\int\, d^D x\sqrt{\gamma}\,\varphi \Big[-\frac{1}{f(n)G(x^a)}\partial^2_t\varphi \\ \nonumber + & G(x^a)\,\partial_n\Big(f(n)\partial_n\varphi\Big)\Big]. \end{align}

There is still a term $G(x^a)$, which is a function of the transverse coordinates in the action, and this term makes it difficult to reduce the theory to effectively two-dimensional in the near horizon limit. To describe the physics near the Killing horizon in terms of the `$t-n$' section of the full spacetime metric in Eq. (\ref{staticmetric}), we impose a constraint on $G(x^a)$ such that the term becomes a constant. To understand this constraint, we calculate the surface gravity $\kappa$ associated with the timelike Killing vector at the Killing horizon $N = 0$. Then the condition  $G(x^a)$ to be constant on the horizon can be mapped into the validity of the zeroth law of black hole mechanics, asserting the constancy of the surface gravity everywhere on the Killing horizon. Therefore, to satisfy the zeroth law, we impose that the lapse function $N(n, x^a)$ is independent of the transverse coordinates in the near horizon limit.\\

To justify this constraint, we note that the zeroth law is an identity for static black hole spacetime, independent of the dynamics of gravity \cite{Racz:1995nh}. Therefore, the surface gravity of a static black hole is a constant on the horizon, irrespective of the field equation and the matter content. Alternatively, we can motivate this choice using the results of \cite{Medved:2004ih}, where the same condition is used to argue the finiteness of the curvature scalar on the horizon. In fact, following \cite{Medved:2004ih}, we may express the lapse function as a Taylor expansion around the horizon at $n = 0$ as,
\begin{equation*}
    N(n,x^a)=\kappa\, n+\frac{\kappa_2(x^a)}{3!}n^3+O(n^4),
\end{equation*}
This Taylor expansion ensures the regularity of the curvature scalars on the horizon and justifies our constraint.\\

Proceeding further, we note that it is always possible to expand any arbitrary function on a compact space in terms of spherical harmonics $Y_{lm}(\theta,\phi\cdots)$, which are functions of spherical coordinates $(\theta, \phi\cdots)$. We can also use the same coordinates in our case. But, to keep the things general, we consider $x^a$ to be a general set of transverse coordinates and use a transformation between $(\theta,\phi\cdots)$ and $x^a$, which allows us to replace ${Y}_{lm}(\theta,\phi\cdots)$ by $\Tilde{Y}_{lm}(x^a)$. Furthermore, $\Tilde{Y}_{lm}(x^a)$ must satisfy a certain normalization condition

\begin{equation}\label{normalization_condition}
    \int dx\sqrt{\gamma}\, \Tilde{Y}_{lm}(x^a)\Tilde{Y}_{l'm'}(x^a)=\delta_{ll'}\delta_{mm'} ,
\end{equation}

Now we can express $\varphi(x^\mu)$ as,
$$\varphi(x^\mu)=\sum_{l,m}\Phi_{lm}(t,n)\Tilde{Y}_{lm}(x^a).$$
If we substitute the above expression of $\varphi$ in Eq. (\ref{actionNH}), the action will have the following form

\begin{align*}
S[\varphi]&=\frac{1}{2}\int dt\, dn \sum_{l,m,l',m'}\Phi_{l'm'} \Big[\Big\{-\frac{1}{N}\partial^2_t\Phi_{lm}\\&+\partial_n(N\partial_n\Phi_{lm})\Big\}  \Big\{\int dx \sqrt{\gamma}\, \Tilde{Y}_{lm}(x^a)\Tilde{Y}_{l'm'}(x^a)\Big\}\Big].
\end{align*}

Using the orthogonality relationship, we then arrive at an expression for the action of the scalar field as,
\begin{align}
S[\varphi]=\frac{1}{2}\int dt\, dn \sum_{l,m}\Phi_{lm} \Big[-\frac{1}{N}\partial^2_t\Phi_{lm}+\partial_n(N\partial_n\Phi_{lm})\Big],
\end{align}
where the lapse function is a function of coordinate $n$ only. Thus, even without the spherical symmetry, our D- dimensional action in the near horizon limit reduces to an action of an infinite set of the scalar fields on the 2-dimensional metric,
\begin{align}\label{2d_metric}
ds^2=-Ndt^2+\frac{1}{N}dn^2.
\end{align}

Given this setting, we can now use the expression of gravitational anomaly as in the case of a spherically symmetric black hole. We discard the ingoing modes that are close to the horizon at $N = 0$, as they have no impact on the behavior of the scalar fields beyond the horizon. As a result, our two-dimensional theory becomes chiral, and the energy-momentum tensor will display an anomaly of the form (\ref{anomaly}) and can be written as,
\begin{align} \label{anomaly_NH}
\nabla_\mu T^\mu_{(\chi)\nu}=A_\nu=\frac{1}{\sqrt{-g}}\,\partial_\mu N^\mu_\nu,
\end{align}
with nonzero components of $N^\mu_\nu$ are 
\begin{align*}
N^n_t&=\frac{1}{192\pi}\Big[\Big(\frac{\partial N}{\partial n}\Big)^2+N\frac{\partial^2 N}{dn^2}\Big], \\
N^t_n&=-\frac{1}{192\pi N^2}\Big[\Big(\frac{\partial N}{\partial n}\Big)^2-\frac{\partial^2 N}{dn^2}N\Big].
\end{align*}
The anomaly described by Eq.(\ref{anomaly_NH}) is timelike in nature as $A_t\neq0$ and $A_n=0$ with the background metric of the form (\ref{2d_metric}). Following the same procedure as in \cite{Robinson:2005pd}, an effective action due to the interaction between metric $g_{\mu\nu}$ and matter can be written as
\begin{align*}
    W[g_{\mu\nu}]=-i\,\ln\Big(\int D[\text{matter}]e^{iS[\text{matter},g_{\mu\nu}]}\Big),
\end{align*}
where $S[\text{matter},g_{\mu\nu}]$ is the classical action functional. Under infinitesimal general coordinate transformation $x^\mu \rightarrow x^\mu -\lambda^\mu$, effective action $W$ changes by

\begin{align}\label{variation}
 \nonumber   -\delta_\lambda W &=\int d^2x \sqrt{-g}\,\lambda^\nu\, \nabla_\mu \Big[T^\mu_{(\chi) \nu}H+T^\mu_{(o)\nu}\Theta_+ \Big]\\ \nonumber &=\int d^2x\, \lambda^t\,\Big[\partial_n\Big(N^n_tH \Big)+  
\Big(T^n_{(o)t}-T^n_{(\chi) t}+N^n_t\Big)\\&\partial \Theta_+ \Big]+\int d^2x \, \lambda^n\,\Big[\Big(T^n_{(o)n}-T^n_{(\chi) n}\Big)\,\partial \Theta_+ \Big].
\end{align}
The function $\Theta_+=\Theta_+(n-\epsilon)$ is a scalar step function, and $H=1-\Theta_+ $ is a scalar ``top hat" function which is 1 in the region $0<n<\epsilon$ and zero elsewhere. Note the integration measure $d^2 x$ is the infinitesimal area element of the $(t-n)$ spacetime. Energy-momentum tensor $T^\mu_{(o)\nu}$ is covariantly conserved in the region $n>\epsilon$. However, the energy-momentum tensor $T^\mu_{(\chi) \nu}$ describes the chiral anomaly through Eq.(\ref{anomaly_NH}). Time independence and Eq.(\ref{anomaly_NH}) limit the possible form of the energy-momentum tensor  $T^\mu_{ \nu}$ up to an arbitrary function of $n$. An Integration over Eq.(\ref{anomaly_NH}) yields the explicit expression for the energy-momentum tensor 
\begin{align*}
   T^t_t&=-\frac{K+Q}{N}-\frac{B(n)}{N}-\frac{I(n)}{N}+T^\alpha_\alpha,\\
   T^n_n&=\frac{K+Q}{N}+\frac{B(n)}{N}+\frac{I(n)}{N},\\
   T^n_t&=-K+C(n)=-N^2T^t_n.
\end{align*}
where  $C(n)=\int_{ 0} ^n\ A_t(n)\,dn$,  
$B(n)=\int_{ 0} ^n\ N A_n(n)\,dn$,  
$ I(n)=\frac{1}{2}\int_{ 0} ^n\ T^\alpha_\alpha \frac{\partial N}{ \partial n}\,dn$, and $K$ and $Q$ are constants of integration. We have taken an assumption that $\frac{I}{N}|_{n=0}=\frac{1}{2}T^\alpha_\alpha|_{n=0}$ to be finite. Note that all terms in the above expression associated with $A_\nu$ vanish in the limit $n\rightarrow 0$. Thus, the variation (\ref{variation}) becomes

\begin{align}\label{variation_NH}
\nonumber -\delta_\lambda W&=\int d^2x\, \lambda^t\,\Big[\partial_n\Big(N^n_tH\Big)+\Big\{-K_o+K_\chi+N^n_t\Big\}\\ \delta(n)\Big] &+\int d^2x \,\lambda^n\,\Big\{\frac{K_o+Q_o-K_\chi-Q_\chi}{N}\Big\}\delta(n).
\end{align}

The general covariance of the full quantum theory demands this variation of the effective action under the diffeomorphism must be zero. But, the above equation also suggests the potential loss of general covariance, in theory, arises from the on-horizon values of the energy-momentum tensor. Nevertheless, our arbitrary variational parameters ($\lambda^t$ and $\lambda^n$) are independent; diffeomorphism invariance implies that each curly brackets term must equal zero, but only on the Killing horizon. This leads to,
\begin{align*}
    K_o&=K_\chi+N^n_t|_{n=0},\\
    Q_o&=Q_\chi-N^n_t|_{n=0},
\end{align*}
where $N^n_t|_{n=0}=(\kappa^2/192\pi)$. We can neglect the finite trace terms as it makes no contribution compared to the divergent $K+Q$ terms in the Killing horizon limit. Thus,
the total energy-momentum tensor $T^\mu_\nu=T^\mu_{(\chi) \nu}H+T^\mu_{(o)\nu}\theta_+ $ transforms into, in the limit $\epsilon\rightarrow0$,
\begin{align*}
 T^\mu_\nu=T^\mu_{\phi \nu}+T^\mu_{c\nu},
\end{align*}
where $T^\mu_{c\nu}$ is our conserved energy-momentum tensor with no quantum effects, and $T^\mu_{c\nu}$ is a conserved tensor with $K=-Q=N^n_t|_{n=0}$, a pure flux. The flux of a massless blackbody radiation beam, which travels in the positive $n$ direction and has a temperature of $T$, has the expression $\phi_s=\pi/(12T^2)$. We require this flux to cancel the gravitational anomaly at the
horizon. Thus comparing $\phi_s$ with $N^n_t|_{n=0}$, we get the Hawking temperature of a
static black hole $T=\kappa/4\pi$, where $\kappa$ is the surface gravity of this static spacetime.\\

Note that the calculation of this section is almost similar to the spherical symmetry case. But this is only possible because, in the near horizon limit, the action of the scalar field becomes effectively two-dimensional. This requires that the lapse function is of a form $N = f(n) G(x^a)$, and the dependence of the transverse coordinates drops off in the near horizon limit. Intriguingly, this can also be motivated from the zeroth law of black hole mechanics, i.e. the constancy of the surface gravity on the horizon.

\section{Stationary Black Hole: Geometric Set up}
In this section, we will do a similar analysis, but for stationary, axisymmetric black holes, and understand Hawking radiation as the gravitational anomaly. \\

 In the coordinate system used in \cite{Medved:2004tp}, the line element of a stationary, axially symmetric black hole spacetime is given by,
 \begin{align}\label{stationarym}
   \nonumber ds^2=-\Tilde{N}(n,z)^2dt^2+g_{\phi\phi}(n,z)\{d\phi-\omega(n,z)dt\}^2+dn^2 \hspace{-1cm}\\ +g_{zz}(n,z)\,dz^2,
\end{align}
where $\omega=-(g_{\phi t}/g_{\phi\phi})$ is an angular-rotation parameter. The existence of two Killing vector fields, $(\partial_t)^\mu$ and $(\partial_\phi)^\mu$, in the above line element makes it clear that spacetime has both axial and time-translational symmetries. As discussed in \cite{Medved:2004tp}, there is a Killing horizon at $\Tilde{N} = 0$ and the surface gravity associated with this Killing horizon is defined as,
\begin{align*}
  \kappa \equiv \lim_{n\rightarrow 0}\partial_n\Tilde{N}.  
\end{align*}
The metric in Eq.(\ref{stationarym}) does not represent the most general form of the stationary axisymmetric black hole spacetime. Apart from being stationary and axisymmetric, we have assumed circularity to write down the above metric. The assumption of circularity has simplified the line element further and allowed us to write the metric with only a single off-diagonal term $g_{t \phi}$. This seems to be a restrictive assumption because there is no guarantee that all stationary, axisymmetric spacetimes to be circular unless it is a vacuum solution of general relativity \cite{Wald:1984rg}. But, recently, it has been proven that even for modified gravity theories, which differ from general relativity perturbatively, all stationary and axisymmetric black hole solutions must be circular \cite{Xie:2021bur}. This result supports our choice of the line element. Also, for a circular black hole, the zeroth law holds independent of the field equations, and the surface gravity is a constant on the horizon \cite{Heus}. Moreover, it has also been established that if the gravity theory is perturbatively related to general relativity, the rigidity theorem holds \cite{Hollands:2022ajj}. All these results provide enough justification to use the line element in Eq.(\ref{stationarym}) for our analysis. This metric represents a stationary and axisymmetric black hole solution of any theory which differs from general relativity perturbatively. \\

The action of a minimally coupled real massless scalar field in this stationary background can be written as,
\begin{align}
\nonumber   
S[\varphi]&=\frac{1}{2} \int d^4x \sqrt{g_{zz}g_{\phi\phi}}\,\varphi\Big[-\frac{1}{\Tilde{N}}\partial^2_t\varphi-\frac{2\omega}{\Tilde{N}}\partial_t\partial_\phi\varphi\\&+\Big(-\frac{\omega^2}{\Tilde{N}}+\frac{\Tilde{N}}{g^{\phi\phi}}\Big)\partial^2_\phi\varphi+\Tilde{N}\partial^2_n\varphi+\frac{\Tilde{N}}{g_{zz}}\partial^2_z\varphi \nonumber\\&+\frac{1}{\sqrt{g_{zz}g_{\phi\phi}}}\Big\{\partial_n\Big(g^{nn}\sqrt{-g}\Big) \partial_n\varphi+\partial_z\Big(g^{zz}\sqrt{-g}\Big) \partial_z\varphi\Big\}\Big], 
\end{align}
where $g=-\Tilde{N}^2g_{\phi\phi}g_{zz}$ is the metric determinant. As in the case of a static spacetime, we assume the lapse function, near the Killing horizon, to be of the form $N(n,z)=f(n)G(z)$ with the property at $n\rightarrow 0$,  $f(n)=0$. However, the explicit expression of $\Tilde{N}(n,z)$, far away from the horizon, is still unknown. An assumption on $G(z)$ to be constant in the Killing horizon limit makes the effective theory 2-dimensional. This arises from the requirement of the validity of the zeroth law of black hole mechanics such that the surface gravity becomes independent of the $z$ coordinate \cite{Medved:2004tp}. We also want our analysis to be consistent with the rigidity theorem. Thus, the angular-rotation parameter $\omega$ must be independent of $z$ on the Killing horizon and denoted by $\omega_H$. In the limit, $n\rightarrow 0$, metric elements $g_{zz}$ and $g_{\phi\phi}$ become a function of $z$ only, and the action takes the form
\begin{align}\label{stationary_action}
\nonumber S[\varphi]=\frac{1}{2}\int d^4x \sqrt{g_{zz}g_{\phi\phi}}\,\varphi\,\Big[-\frac{1}{\Tilde{N}}\partial^2_t\,\varphi\\-\frac{2\, \omega_H}{\Tilde{N}}\,\partial_t\partial_\phi\varphi-\frac{\omega_H^2}{\Tilde{N}}\,\partial^2_\phi\varphi +\partial_n\Big(\Tilde{N}\partial_n\varphi\Big)\Big], 
\end{align}
 Consider a transformation \cite{Murata:2006pt},
\begin{align}
 \nonumber   \psi&=\phi-\omega_Ht,\\ 
    \xi&=t,
\end{align}
to eliminate $t-\phi$ and $\phi-\phi$ derivative terms from the Eq.(\ref{stationary_action}), and rewrite the action as
\begin{align}
    S[\varphi]=\frac{1}{2}\int d^4x\sqrt{g_{zz}g_{\phi\phi}}\,\varphi \,\Big[-\frac{1}{\Tilde{N}}\partial^2_\xi\varphi+\partial_n\Big(\Tilde{N}\partial_n\varphi\Big)\Big].
\end{align}
Note that coordinate $z$ on the Killing horizon is arbitrary. Therefore, we assume here also exists a co-ordinate transformation (similar to the case of static spacetime discussed in the previous section) under which  $Y_{lm}(\theta,\phi)$ transformed into $\Tilde{Y}_{lm}(\phi,z)$ and $\Tilde{Y}_{lm}(\phi,z)$ satisfies a normalized condition similar to the Eq.(\ref{normalization_condition}).
Decomposing $\varphi(x^\mu)$ into $\Phi_{lm}(\xi,n)$ and $\Tilde{Y}_{lm}(z, \phi)$, we can write down the action in the following simple form,
\begin{align}\label{stationary_action_NH}
S[\varphi]=\frac{1}{2}\int d\xi dn\sum_{l,m}\Phi_{lm}\Big[-\frac{1}{\Tilde{N}}\partial^2_\xi\Phi_{lm}+\partial_n\Big(\Tilde{N}\partial_n\Phi_{lm}\Big)\Big],
\end{align}
where the lapse function $\Tilde{N}$ depends on $n$ only. Thus, The Eq.(\ref{stationary_action_NH}) is an
action for an infinite set of scalar fields in the 2-dimensional spacetime with the metric,
\begin{align}
ds^2=-\Tilde{N}d\xi^2+\frac{1}{\Tilde{N}}dn^2.
    \end{align}
Now using a similar argument explained in Sec. II, we can find the form of the Hawking temperature for the stationary black holes as $T=(\kappa/4\pi)$, where $\kappa$ is the surface gravity of the Killing horizon in this stationary spacetime.

\section{Discussions and conclusion}\label{con}

The derivation of the Hawking radiation using gravitational anomaly depends on the near horizon geometry. Since the gravitational anomaly appears only for $(4k + 2)$ dimensions, the near horizon physics is anomalous, provided it is effectively two-dimensional. In the case of spherically symmetric spacetimes, this can be easily established by integrating over the angular coordinates. Also, for specific solutions beyond spherical symmetry, like a Kerr black hole, this is shown to be true \cite{Murata:2006pt}. These suggest that there must be a general derivation that can establish the effective two-dimensional nature of the physics near a general stationary black hole horizon. In our work, we have demonstrated such property and derived of the Hawking radiation flux from gravitational anomaly for a general stationary black hole.

Our work suggests an intriguing feature that the derivation works provided the surface gravity is a constant on the horizon i.e. the zeroth law of black hole mechanics holds. The zeroth law has been first established for stationary black hole solutions of general relativity using dominant energy condition \cite{Wald:1984rg}. But, later, the derivation was generalized for the Lovelock class of theories \cite{Ghosh:2020dkk}. Recently, it was shown that the zeroth law holds for any metric theory of gravity provided it is perturbatively connected to general relativity \cite{Bhattacharyya:2022nqa}. Therefore, we are justified to impose the constancy of the surface gravity on the black hole horizon in our geometric setup, and that leads to the derivation of Hawking radiation as a gravitational anomaly. This is also consistent with the idea that Hawking radiation is only dependent on the geometric structure of the horizon, independent of the dynamics of gravity \cite{Visser:1997yu}.

\section*{Acknowledgement}
The research of SS is supported by the Department of Science and Technology, Government of India, under the SERB CRG Grant (CRG/2020/004562). SK acknowledge Sabarmati Bridge Fellowship from IIT Gandhinagar for Support. The authors thank Rajes Ghosh for extensive discussion.

\end{document}